\documentclass[12pt]{article}
\usepackage{amsfonts}
\usepackage{graphicx}
\usepackage{amsmath}
\usepackage{cite}

\usepackage{color}
\definecolor{darkviolet}{rgb}{0.58, 0.0, 0.83}

\begin{document}

\title{Continuous variable graph states: entanglement and  graph properties}

\author{Kh. Gnatenko$^{1,2}$\footnote{khrystyna.gnatenko@gmail.com}, V. Tkachuk$^1$\footnote{voltkachuk@gmail.com}, M. Krasnytska$^{3,4}$\footnote{kras.marjana@gmail.com}, Yu. Holovatch $^{3,4,5}$\footnote{hol@icmp.lviv.ua}\\
$^1$Professor Ivan Vakarchuk Department for Theoretical Physics,\\
Ivan Franko National University of Lviv, 79005 Lviv, Ukraine.\\
$^2$ SoftServe Inc., 2d Sadova St., 79021 Lviv,  Ukraine.\\
$^3$ Institute for Condensed Matter Physics,\\ National Academy of Sciences of Ukraine, UA-79011 Lviv, Ukraine.\\
$^4$ $\mathbb{L}^4$ Collaboration  $\&$ Doctoral College  for the Statistical Physics \\ of Complex Systems,
Leipzig-Lorraine-Lviv-Coventry, Europe\\
$^5$ Centre for Fluid and Complex Systems, Coventry University,\\ Coventry CV1 5FB, UK.}

\maketitle

\begin{abstract}
We propose the definition of the geometric
measure of entanglement for continuous variable states. On the basis of this definition we examine entanglement of the graph states  obtained as a result of
action of a unitary operator on the ground state of a system of $N$ noninteracting harmonic oscillators.
We  find that the entanglement of a harmonic oscillator with other ones
is defined by the value of its vertex degree.
\end{abstract}

\section{Introduction}\label{I}

Entanglement is one of the most intriguing features of quantum physics. It has been described in 30-ies of
the last century in classical works of Einstein, Podolsky, Rosen, and Shr\"odinger \cite{Ein35,Schrod35}
however its intensive studies are carried out in the last  two decades (see, for instance, \cite{Hor09, Shi95,Behera,Scott,Horodecki1,Ple07,Torrico,Sheng,Samar,Kuzmak,Gnatenko,Kuzmak2,Susulovska,Wang,Mooney}).
This recent progress to the large extent it is due to the role the entanglement plays in various
quantum-information processing, quantum cryptography, quantum teleportation
\cite{Hor09,Bennett,Bouwmeester,Ekert,Raussendorf,Lloyd,Buluta,Shi,Llewellyn,Huang,Yin,Jennewein,Karlsson}.
In the context of this paper let us mention also that the
entanglement of  two coupled quantum harmonic oscillators has been examined in  \cite{Makarow} whereas
the entanglement dynamics of coupled harmonic oscillators was studied in \cite{Plenio2004}.

One of the most natural entanglement measures is the geometric measure, which was proposed by Shimony
\cite{Shi95} and generalized for multipartite system by Wei and Goldbart \cite {Wei03}. This measure is
defined as a minimal squared distance between an entangled state $| \psi\rangle$ and a set of
separable states $|\psi_s\rangle$
\begin{eqnarray} \label{DefE}
E=\min_{|\psi_s\rangle}(1-|\langle\psi|\psi_s\rangle|^2)=1-\max_{|\psi_s\rangle}|\langle\psi|\psi_s\rangle|^2,
\end{eqnarray}
where $1-|\langle\psi|\psi_s\rangle|^2$ is the so-called squared Fubini-Study distance.
Next we will use the notation
\begin{equation} \label{2}
I=\langle\psi|\psi_s\rangle \, .
\end{equation}

Despite its simple definition, the measure (\ref{DefE}) involves minimization
procedure over separable states. The last is non-trivial and therefore an
explicit value of entanglement geometric measure was derived only for the limited
number of entangled states such as GHZ states \cite{Wei03}, Dicke
states \cite{Wei03,Mar10}, generalized W states \cite{Tam10},
graph states \cite{Mar07} and other type of symmetric states (see
also papers \cite{Tam08,Sai08,Hay09,Tam102,Chen10,Str11}).
Recently, in paper \cite{Samar} the relation of the geometric measure of entanglement to
mean values of an observable of the entangled system was found.
The authors showed that for pure states
the geometric measure of entanglement of a spin  with arbitrary
quantum system  is determined by its mean value. In turn, for mixed rank-2
states the entanglement is related to the values of spin correlations.
These results simplify the procedure of quantification of entanglement.

In the present paper the way of detection of   the geometric
measure of  entanglement of continuous variable states is found. The geometric measure of entanglement of the graph states  obtained as a result of
action of a unitary operator on the ground state of a system of $N$ noninteracting harmonic oscillators is examined.
We show that the  entanglement of a harmonic oscillator with other ones in the system  depends on the value of the vertex degree in the corresponding graph.

The paper is organized as follows. In Section \ref{II}
the  geometric measure of an entanglement in continuous variable system is examined.
Section \ref{III} is devoted to studies of the geometric measure of an entanglement of graph states.
In particular, we consider a graph state that results from an action of the unitary
operator on the ground state of a system of noninteracting harmonic oscillators.
We find the relation between the entanglement of a single harmonic oscillator with properties
of the graph that defines the unitary operator. We end by conclusions in Section \ref{V}.

\section{Geometric measure of entanglement in continuous variable system}\label{II}
Let us consider a quantum system which consists of two subsystems living in the configurational space with
coordinates ${\bf x}_1=(x_1^1,x_1^2,...,x_1^n)$ and  ${\bf x}_2=(x_2^1,x_2^2,...,x_2^m)$ for the first and
second subsystems, respectively. The quantum state of the whole system is described by the wave function
\begin{eqnarray}\label{psie}
\psi=\psi({\bf x}_1,{\bf x}_2),
\end{eqnarray}
which is normalized
\begin{eqnarray}
\int dV_1dV_2|\psi({\bf x}_1,{\bf x}_2)|^2=1.
\end{eqnarray}

When the wave function (\ref{psie}) can be written as
\begin{eqnarray}
\psi_s=\phi_1({\bf x}_1)\phi_2({\bf x}_2),
\end{eqnarray}
where $\phi_1({\bf x}_1)$, $\phi_1({\bf x}_1)$ are the wave functions for the first and second subsystems,  
the corresponding quantum state is called factorized and the measure of the entanglement is zero. The 
wave functions for the first and second subsystems are normalized
\begin{eqnarray}\label{Norm}
\int dV_1|\phi_1({\bf x}_1)|^2=1, \ \ \int dV_2|\phi_2({\bf x}_2)|^2=1\, .
\end{eqnarray}

In order to find the rate of entanglement in state (\ref{psie}) in a general case, let us use the
geometric measure of entanglement (\ref{DefE}). Substituting (\ref{psie}) into (\ref{2}) readily leads to
\begin{eqnarray}
I=\int dV_1dV_2\psi^*({\bf x}_1,{\bf x}_2)\phi_1({\bf x}_1)\phi_2({\bf x}_2).
\end{eqnarray}
To proceed with (\ref{DefE}), it is necessary to find the maximal value of this functional, ${\rm max} |I|^2$,
with respect to $\phi_1({\bf x}_1)$ and $\phi_2({\bf x}_2)$ taking into account normalization conditions
(\ref{Norm}). For this purpose let us consider the functional
\begin{eqnarray}
F[\phi_1,\phi_2]=|I|^2-\lambda_1\left(\int d V_1 |\phi_1({\bf x}_1)|^2-1\right)-\lambda_2\left(\int d V_2 |\phi_2({\bf x}_2)|^2-1\right),
\end{eqnarray}
where $\lambda_1$ and $\lambda_2$ are Lagrange multipliers.
Maximum of the functional $F$ can be found considering the equation for its variation $\delta F=0$ which gives
\begin{eqnarray}
{\delta F\over \delta \phi_1({\bf x}_1)}=0, \ \ {\delta F\over \delta \phi_2({\bf x}_2)}=0,
\end{eqnarray}
or complex conjugated equations
\begin{eqnarray}
{\delta F\over \delta \phi^*_1({\bf x}_1)}=0, \ \ {\delta F\over \delta \phi^*_2({\bf x}_2)}=0.
\end{eqnarray}
In an explicit form the last equations read
\begin{eqnarray}\label{Eigen1}
I\int dV_2\psi({\bf x}_1,{\bf x}_2)\phi^*_2({\bf x}_2)=\lambda_1 \phi_1({\bf x}_1),\\  \label{Eigen2}
I\int dV_1\psi({\bf x}_1,{\bf x}_2)\phi^*_1({\bf x}_1)=\lambda_2 \phi_2({\bf x}_2).
\end{eqnarray}
Multiplying the first equation by $ \phi^*_1({\bf x}_1)$ and integrating over ${\bf x}_1$, and multiplying the second equation by
$\phi^*_2({\bf x}_2)$ and integrating over ${\bf x}_2$ we arrive at
\begin{eqnarray}
\lambda_1=\lambda_2=\lambda=|I|^2.
\end{eqnarray}

Substituting $\phi_2({\bf x}_2)$ from (\ref{Eigen2}) into (\ref{Eigen1}) we obtain
\begin{eqnarray}
\int dV_2dV'_1\psi({\bf x}_1,{\bf x}_2) \psi^*({\bf x'}_1,{\bf x}_2)\phi_1({\bf x'}_1)=\lambda \phi_1({\bf x}_1) \, .
\end{eqnarray}
This equation can be rewritten in the form
\begin{eqnarray}\label{EigenF}
\int dV'_1 K({\bf x}_1,{\bf x'}_1) \phi_1({\bf x'}_1)=\lambda \phi_1({\bf x}_1) \, ,
\end{eqnarray}
where the integral kernel function
\begin{eqnarray}\label{k}
 K({\bf x}_1,{\bf x'}_1)=\int dV_2\psi({\bf x}_1,{\bf x}_2) \psi^*({\bf x'}_1,{\bf x}_2)
\end{eqnarray}
is the density matrix of the first subsystem. Equation (\ref{EigenF}) can be treated as an eigenvalue equation where eigenvalue $\lambda=|I|^2$.
Thus $|I|_{\rm max}^2=\lambda_{\rm max}$ and the geometric measure of entanglement reads
\begin{eqnarray}\label{ent}
E=1-\lambda_{\rm max} \label{E} \, .
\end{eqnarray}
Therefore, the problem of calculation of the geometric measure of entanglement has been reduced to defining the eigenvalue $\lambda$  of equation (\ref{EigenF}).

\section{A case study: harmonic oscillators graph state}\label{III}

With the geometric measure to quantify entanglement at hand, Eqs. (\ref{EigenF}), (\ref{ent}), it is
straightforward to proceed implementing it to particular cases of interest. 
To this end, let us
consider the  graph state. Such  states are widely studied because of its importance in quantum 
information and quantum computing (see, for instance, \cite{Pfister,Gu}).

\begin{figure}
\begin{center}
 \includegraphics[width=6cm]{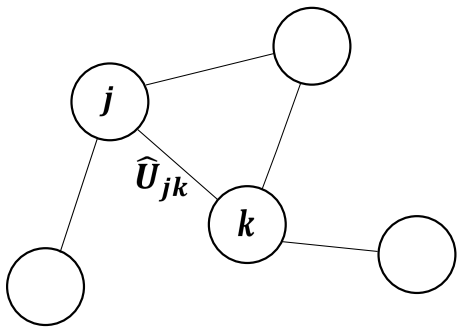}
 \caption{Harmonic oscillator graph state: each oscillator
can be associated with the graph vertex, action of the unitary operator  $\hat{U}_{jk}$
on the ground state of a system of $N$ non-interacting harmonic oscillators  corresponds to the edge
between the vertices $j$, $k$. \label{fig1}
}
\end{center}
\end{figure}

As a case study of quantifying entanglement for the continuous variable quantum state,
let us consider the graph state defined as
\begin{eqnarray}\label{4}
\psi({\bf x})=\psi(x_1,x_2,...,x_N)=(\alpha/\pi)^{N/2}\exp\left({-\sum_{j}\frac{\alpha x_j^2}{2}+i\sum_{j,k}\frac{a_{jk}x_j x_k}{2}}\right),\label{s}
\end{eqnarray}
here $\alpha$ is a constant, $a_{jk}$ are elements of a constant symmetric matrix $\hat{a}$, indexes $j,k=(1, \dots ,N)$.
To associate this state with a graph of $N$ vertices we assume that matrix  $\hat{a}$ is given by the graph adjacency matrix with components $a_{jk}=1$
if there is an edge between vertices $j$ and $k$ and  $a_{jk}=0$ otherwise, $a_{jj}=0$.
State (\ref{s}) can be obtained as a result of an action of the unitary operator
 \begin{eqnarray}
 \hat{U}_{jk}=e^{ia_{jk}x_jx_k}\label{u}
  \end{eqnarray}
on the ground state of a system of $N$ non-interacting harmonic oscillators $\psi_0({\bf x})=(\alpha/\pi)^{N/2}\exp(-\sum_{j}\alpha x_j^2)$.
Each oscillator can be associated with graph vertex whereas the terms $i\sum_{j,k}{a_{jk}}x_jx_k/2$ in the exponent (\ref{s}) correspond
to the edges between the vertexes and appear as a result of action of the operator $\hat{U}_{jk}$ (\ref{u}), see Fig. \ref{fig1}.
The wave function (\ref{4}) describes a system
of interacting ``kicked'' harmonic oscillators (see, for instance, \cite{Gardiner}).

Let us consider an entanglement of one oscillator with coordinate $x_1$ with the others.
In this case ${\bf x}_1=(x_1)$ and
${\bf x}_2=(x_2,x_3,x_4...,x_N)$. As it was shown in the previous section, the geometric measure of an entanglement of a state  is determined by the maximal
eigenvalue of equation (\ref{EigenF}). Taking into account (\ref{k}) in the case when the state is defined as (\ref{s}) one has
\begin{eqnarray}\label{EigenF1}
 K({x}_1,x'_1)=\int dx_2dx_3...dx_N\psi^{\ast}(x_1,x_2,...,x_N)\psi(x'_1,x_2,...,x_N)=\nonumber\\=\sqrt{\frac{\pi}{\alpha}}
 \exp\Big (-\frac{\kappa}{4\alpha}(x_1-x'_1)^2-\frac{\alpha}{2}(x_1^2+(x'_1)^2)\Big ) \, ,
\end{eqnarray}
where
\begin{equation}
\kappa_1=\sum_ja_{1j}\, .\label{kappa}
\end{equation}
So, to find the geometric measure of an entanglement one has to consider the following equation
\begin{eqnarray}
\sqrt{\frac{\pi}{\alpha}}\int dx'_1\exp\left(-\left(\frac{\kappa_1}{4\alpha}+\frac{\alpha}{2}\right)(x_1^2+(x'_1)^2)+\frac{\kappa_1}{2\alpha}x_1x'_1\right) \phi_1(x'_1)=\lambda \phi_1({ x}_1).\label{e4}
\end{eqnarray}
The solution of (\ref{e4}) reads
\begin{eqnarray}
\phi_n(x_1)=\sum^n_{i=0}C_i x_1^i e^{-\frac{\kappa_1}{2\alpha}x_1^2},\\
\lambda_n=\frac{2\alpha \kappa_1^n}{(\kappa_1+2\alpha^2+2\alpha\sqrt{\alpha^2+\kappa_1})^{n+\frac{1}{2}}},\label{egn}
\end{eqnarray}
with $C_i$ being constants. It is easy to show that the spectrum of eigenvalues $\lambda_n$ attains maximal value at $n=0$ leading to
\begin{eqnarray}
\lambda_{\rm max}=\lambda_0=\frac{2\alpha}{\sqrt{\kappa_1+2\alpha^2+2\alpha\sqrt{\alpha^2+\kappa_1}}}=
\frac{2}{1+\sqrt{1+\kappa_1/\alpha}}\label{max} \, .
\end{eqnarray}
Finally, taking into account (\ref{ent}) one gets that the geometric measure of an entanglement of a single oscillator
in interacting kicked oscillator graph state is determined as
\begin{eqnarray}
E=\frac{\sqrt{1+\kappa_1/\alpha}-1}{\sqrt{1+\kappa_1/\alpha}+1}.\label{ent1}
\end{eqnarray}
It is worth noting that $\kappa_1$ as given by (\ref{kappa}) is nothing else but the  degree of the vertex 1 that
corresponds to the oscillator whose entanglement is being measured. Therefore, according to (\ref{ent1}) an
entanglement of a harmonic oscillator depends on the value of degree of the vertex which represents it.
Note that the entanglement  increases with an increase of $\kappa_1$. This manifests correlation of an entanglement
of a harmonic oscillator with the number of edges
incident to the corresponding vertex  in the graph.

\section{Conclusions}\label{V}

We propose the way to quantify the geometric measure of entanglement for continuous variable states. We show
that the entanglement is related with the maximal eigenvalue of the eigenvalue equation (\ref{EigenF}).

On the basis of the proposed way to quantify  entanglement (\ref{E}) we have studied entanglement of graph
states. We have considered the states  obtained as a result of an action of the unitary operator
$\hat{U}_{jk}=\exp(ia_{jk}x_jx_k)$ on the ground state of a system of $N$ noninteracting harmonic oscillators.
We have obtained that the entanglement of a single harmonic oscillator depends on the value of degree of the vertex
which represents it. So, there is a correlation of the entanglement of a harmonic oscillator with number of edges
incident to the corresponding vertex  in the graph.

\section*{Acknowledgments}
Kh.G. and V.T. acknowledge support from National Research Foundation of Ukraine,
Project 2020.02/0196  (No. 0120U104801).

\end{document}